\documentstyle[12pt]{article}
\newcommand{\beq}{\begin{equation}}
\newcommand{\eeq}{\end{equation}}
\newcommand{\rr}{I\!I}

\begin{document}
\pagestyle{empty}
\begin{flushright}
{BROWN-HET-1040} \\
\end{flushright}
\vspace*{5mm}

\begin{center}
{\Large{\bf Non-abelian Eikonals}}\\
[10mm]

H. M. Fried\footnote{Supported in part by DOE Grant DE-FG02-91ER40688 - Task
A}\\
{\em Department of Physics \\
Brown University \\
Providence RI 02912 USA}\\
[5mm]
and\\
[5mm]
Y. Gabellini\\
{\em Institut Non-Lin\'eaire de Nice, 1361 Route des Lucioles\\
06560 Valbonne France}\\
[2cm]
Abstract
\end{center}
\vspace*{3mm}
A functional formulation and partial solution is given of the non-abelian
eikonal problem associated with the exchange of non-interacting, charged or
colored bosons between a pair of fermions, in the large $s$/small $t$ limit.  A
simple, functional  ``contiguity" prescription is devised for extracting those
terms which exponentiate, and appear to generate the leading, high-energy
behavior of each perturbative order of this simplest non-abelian eikonal
function; the lowest non-trivial order agrees with the corresponding SU(N)
perturbative amplitude, while higher-order contributions to this eikonal
generate an  ``effective Reggeization" of the exchanged bosons, resembling
previous results for the perturbative amplitude.  One exact and several
approximate examples are given, including an application to self-energy
radiative corrections. In particular, for this class of graphs and to all
orders in the coupling, we calculate the leading-log eikonal for SU(2).  Based
on this result, we conjecture the form of the eikonal scattering amplitude for SU(N).

\newpage
\setcounter{page}{1}
\pagestyle{plain}

\section{Introduction}
\indent

One of the most persistent problems in the application of field theory methods
to particle scattering has been the inability to generalize, in a direct
functional, non-perturbative way, abelian eikonal models to their non-abelian
counterparts.  Many efforts in this direction have of course been made over the
last several decades, using the partial, perturbative summation of an eikonal
function \cite{C-W}, or a variety of non-perturbative approximations \cite{HMF
77}. In ref. \cite{HMF 77}, for example, a ``mean-field" approximation was made
to the relevant functional integrals corresponding to the exchange of neutral
vector mesons (NVMs) between scattering nucleons, which include the
restrictions of SU(2) isospin; and the result, while ``approximately correct",
left a certain unease in its wake.  A more modern example is the problem of how
to include SU(3) color restrictions in QCD$_4$, which must be faced if one is
to attempt any functional calculation using the recent, exact and approximate
Greens' functio
ns $G_c(x,y|A)$ of that theory \cite {HMF-YG 1}; or, indeed, the new
dimensional-transmutation/flux-string expansion of quark-quark scattering
amplitudes \cite{Blois}.

We give in this paper a complete, if formal, representation of the simplest
non-abelian eikonal, corresponding to multiple gluon exchange between
scattering quarks {\bf without} virtual gluon-gluon interactions; we extract
that portion which can be easily isolated, and define a particular,
ordered-exponential (OE) representation of the remainder, which can be expanded
or approximated in various ways. In particular, we define a simple, functional
procedure called ``contiguity", which, in an immediate way, isolates at least a
subset of those terms that are definitely exponentiated, and can be represented
to all orders by a perturbative expansion of the eikonal.  These terms
correspond to the leading $s$-dependence in the lowest, non-trivial order, and
we argue that they correspond to the extraction of the leading $s$-dependence
in every perturbtive order of the eikonal function.  For quark-quark
scattering, the result duplicates the essence of well-known, leading-log
perturbative estimates previously calculate
d for amplitudes \cite{C-W}. The method will be illustrated in two contexts,
and its applicability discussed for more general, non-abelian problems; in
particular, for this class of graphs and to all orders in the coupling, we
calculate the leading-log eikonal of SU(2).

To our knowledge this is the first time that such estimates have been obtained
in a purely functional context, while the contiguity technique opens the way
for an attack on other, more complicated, non-abelian eikonal problems, such as
those which involve virtual gluon-gluon interactions (in particular, the
so-called ``towers" and their generalizations), as well as self-energy and
vertex effects of non-abelian, virtual-gluon emission and absorption by a
single quark.
 However, by treating only boson exchange, without self-interactions between
the exchanged bosons, we are apparently going to violate requirements of gauge
invariance, which for perturbative, Yang-Mills gluons, require the simultaneous
computation of all relevant graphs of a given order, and not just the simple
eikonal graphs considered here.  Surely the same sort of inclusion must
eventually be true for any non-perturbative attempt.  We ask the reader to
suspend judgment on this point until the final discussion presented in the
Summary of Section 6; and to realize that, while gauge invariance must of
course be insured in any computation whose results are going to be compared
with experiment, we are proposing a functional attach on that part of the
problem of immediate concern to the scattering quarks.  This is important
because a functional treatment contains all powers of the coupling; and it is
useful because there exists an additional, computational step by which gauge
invariance can be re-established - including the relevant contributions
generated by all gluon-gluon interactions - later on.  The main thrust of the
present paper is the functional extraction of leading-log, energy dependence of
the simplest, non-Abelian eikonal.

We begin at that stage of a quark-quark scattering amplitude where mass-shell
amputation (MSA) has already been carried out on the fermion Green's functions
$\langle p_{1,2} \vert G_c [A] \vert p_{1,2}^{\prime}\rangle $ approximated in
a no-recoil fashion \cite{HMF B1}, and the essential structure of the eikonal
function which describes non-abelian NVM exchange between a pair of fermions
(quarks, for SU(3)) has been recognized \cite{HMF B2} as:

\begin{eqnarray}
e^{\displaystyle{i\chi}} &=& e^{\displaystyle{-i\int {\delta\over\delta A_I}
\,Q\, {\delta\over\delta A_{\rr}}}} \left( e^{\displaystyle{-ig_1
\int_{-\infty}^{+\infty} ds p_{1}^{\mu} A_{I\mu}^{a} ( z_{1} - s p_{1} )
\lambda_{a}^{I} }}\right)_+\nonumber\\
&& \left( e^{\displaystyle{-ig_{2} \int_{-\infty}^{+\infty} dt p_{2}^{\nu}
A_{\rr\nu}^{b} ( z_{2} - tp_{2} ) \lambda_{b}^{\rr}
}}\right)_{+{\textstyle\left|{A_{I} = A_{\rr} = 0, g_{1} = g_{2} = g
}\right.}}
\end{eqnarray}
where $z_{1,2}$ and $p_{1,2}$ are the fermions' configuration and momentum
coordinates, and $Q_{\mu\nu}^{ab}$ is the appropriate boson propagator. Eq.
(1.1) defines ``linkages" between a pair of OEs, and the result will
necessarily be a ``doubly-ordered-exponential".  How this can be transformed
into a pair of single OEs; how the leading-logs of the latter may be extracted,
leaving but a single OE; and how that OE can, for SU(2), be summed explicitly
over all perturbative orders, is the main content of this paper.
\par
More precisely, the preferred method of obtaining the eikonal in the
conventional case, where the conventional, no-recoil approximation of $G_c[A]$
destroys coordinate symmetry of this Green's function, is to calculate not
$T_{eik}$ but, before MSA:

\begin{eqnarray}
&{\displaystyle{\partial^2 T_{eik}\over \partial g_1 \partial g_2}}&  =
{i\over  g_1 g_2} \, {\delta\over\delta\phi (0)} \, {\delta\over \delta\psi(0)}
\,\, e^{\displaystyle{-i\int {\delta\over \delta A_{I}} Q {\delta\over\delta
A_{\rr}}}}\nonumber \\
& & \left( e^{\displaystyle{-ig_{1}\int_{-\infty}^{+\infty} ds \phi (s)
p_{1}\cdot A_{I} (z_{1} - sp_{1} )\cdot \lambda^{I}}} \right)_+\\
& & \left( e^{\displaystyle{-ig_{2} \int_{-\infty}^{+\infty} dt \psi (t)
p_{2}\cdot A_{\rr} (z_{2} - tp_{2} ) \cdot \lambda^{\rr}}} \right)_{+
{\textstyle\left|{\phi (s)=\psi (s) = 1, A_{I} = A_{\rr} = 0}\right.
}}\nonumber
\end{eqnarray}
and integrate over $g_{1,2}$ (with the boundary conditions $T_{eik} (g_1 , 0) =
T_{eik} (0,g_2 ) = 0$) after the necessary functional linkages have been
performed \cite{HMF B2}; it has been assumed that the RHS of (1.2) is a
function of $z_1 - z_2$, and the subsequent $\delta^{(4)} (q_1 + q_2 )$
statement of 4-momentum conservation has been suppressed.  For simplicity we
consider the quantity of (1.1) as representative of the correct eikonal -- it
is exactly the eikonal in the absence of non-abelian complications -- even
though it is quite possible to produce, upon integration over the couplings of
(1.2), combinations which are more complicated than that of (1.1).  However,
eq. (1.1) is representative of the full, non-abelian structure of the problem,
and we here restrict attention to this quantity.  The non-commuting objects
$\lambda_a$ are taken to be the Gell-Mann matrices of SU(N).  We again
emphasize that more complicated eikonal graphs, such as the "tower graphs" of
Cheng and Wu \cite{C-W}, are not inc
luded in this analysis, although they can be formally inserted by the
functional methods outlined in the last chapters of references \cite {C-W} and
\cite{HMF B1}.

In the abelian case, where $A_{\mu}^a \rightarrow A_{\mu}$, and the $\lambda_a$
are missing, the functional operation of (1.1) may be performed immediately,
yielding:

\beq
i\chi = ig^2 (p_1 \cdot p_2 ) \int\!\!\int_{-\infty}^{+\infty} ds \, dt
\Delta_c \left( z_1 - z_2 - sp_1 + tp_2 \right)
\eeq
with a propagator $Q_{\mu\nu} (x_1 , x_2 )= \delta_{\mu\nu} \Delta_c (x_1 - x_2
)$.  The proper-time integrals are easily performed when a Fourier
representation of $\Delta_c$ is inserted into (1.3); and one finds:

\beq
i\chi = - i {g^2\over 2\pi} \gamma (s) \, K_0 (\mu b )
\eeq
where $\gamma (s) = {\displaystyle{(s-2m^2)\over \sqrt{s ( s-4m^2 )}}}$ is that
factor depending on the spin of the exchanged boson, of mass $\mu$; the fermion
mass is denoted by $m$, and in this equation, $s$ denotes the total CM
(energy)$^2$ of the two quarks.  In all subsequent expressions, we shall assume
the high-energy limit, where $\gamma (s) \rightarrow 1$.
\par
We give in the next Section a new, functional formulation of the eikonal of
(1.1), and, in an appropriate kinematical situation, display one exact
solution.  More generally, a perturbative expansion of this eikonal functional
may be defined, and certain obvious terms (which are the most elementary
generalizations of the abelian eikonal) are summed to all orders.  In Section
III, we define the statement of ``functional contiguity", which isolates those
terms of (1.1) that are definitely exponentiated, and which appears to generate
the leading $\ln (E/m)$ dependence of every perturbative term of the
non-abelian eikonal, when the necessary, doubly-ordered-exponential is defined
in a moderately elegant way.  In the next Section, we discuss the leading-log
approximation, and show how the extraction of such terms (from ``nested"
momentum integrals) can reduce the complexity of the computations to operations
upon a single OE; for SU(2), these operations are performed and summed to all
orders, and suggest a conjectu
re for the corresponding eikonal scattering amplitude of SU(N). In Section V,
we apply the analysis to self-energy processes, as well as to eikonal tower
graphs and their generalizations, while Section VI contains a summary of our
present understanding of this eikonal construction.

\section{Formulation}
\setcounter{equation}{0}
\indent

In order to perform the functional operation of (1.1), it is useful to
introduce for each OE the functional representation:

\begin{eqnarray}
\left( e^{\displaystyle{-ig\int_{-\infty}^{+\infty} ds p_{\mu} A_{\mu}^{a} (z -
sp ) \lambda_a }} \right) _ + = &&\\
  N^{\prime} \int d [\alpha ] \int d[u] \,
e^{\displaystyle{i\int_{-\infty}^{+\infty} ds \alpha_{a} (s) [ u_{a} (s) - g
p_{\mu} A_{\mu}^{a} (z - sp ) ] }}& &\!\!\!\! \cdot \left(
e^{\displaystyle{i\int_{-\infty}^{+\infty} ds \lambda_{a} u_a (s)}} \right) _+
\, \,\nonumber
\end{eqnarray}
or, more simply, rewriting (2.1) as: $I\otimes \exp \left[ - i
\int_{-\infty}^{+\infty} ds \, p_{\mu} A_{\mu}^a (z-sp) \alpha_a (s)\right], $
where $N^{\prime}$ is an appropriate normalization constant.  That (2.1) is
trivially true can be seen by breaking up the $-\infty < s < + \infty$ range
into small intervals, and integrating over the $\alpha_a (s_i )$ which leads to
a delta functional of the $u_a (s)$, whose integration immediately produces the
LHS of (2.1).  The advantage of this procedure is that the functional linkages
of (1.1) are now abelian, and may be performed immediately, yielding:

\beq
e^{i\chi} = I_1 \otimes \cdot I_2 \otimes \exp \left[ i \int\!\int_{-\infty}^{+
\infty} ds \, dt \, \alpha_a (s) Q_{ab} (s,t) \beta_b (t) \right]
\eeq
with $Q_{a,b} = g^2 p_1^{\mu} Q_{\mu\nu}^{ab} p_2^{\mu}$, and where the
$I_{1,2}$ denote, from (2.1), simultaneous functional operations to be
performed on the $\alpha_a (s )$ and $\beta_b (t )$ variables.
\par
These final operations are what is now needed, and may be delineated by the
insertion of relevant source and parameter dependence, followed by a
``Schwingerian search" for an appropriate ``differential characterization".
With the definition:

\begin{eqnarray}
R (s,t\vert \xi , \eta ) & = & N^{\prime} \int d [\alpha ] \int d [u] \,
e^{i\int_{-\infty}^{+\infty} \alpha \cdot u} \left( e^{i\int_{-\infty}^{s}
\lambda^{I} \cdot u} \right) _+ \, e^{i\int_{-\infty}^{+\infty} u \cdot \xi
}\nonumber\\
& & \cdot N^{\prime} \int d [\beta] \int d [v] \,
 e^{i\int_{-\infty}^{+\infty} \beta \cdot v} \left( e^{i\int_{-\infty}^{t}
\lambda^{\rr} \cdot v} \right)_+ \, e^{i\int_{-\infty}^{+\infty} v \cdot \eta}
\\
& & \cdot \exp \left[ i \int\!\int_{-\infty}^{+\infty} ds^{\prime} dt^{\prime}
\alpha_a (s^{\prime} ) Q_{ab} ( s^{\prime} , t^{\prime} ) \beta_b (t^{\prime} )
\right] \nonumber
\end{eqnarray}
comparison with (2.2) shows that the quantity needed is $\ln R (+\infty ,
+\infty \vert 0, 0 )$.  One can create a variety of differential equations
involving the proper-time parameters $s, t$ and the sources $\xi_a (s), \eta_b
(t)$; but for present purposes, it seems to be sufficient to work with only $s$
and $\eta$, so that we consider $R (s,+\infty \vert 0, \eta ) = R(s \vert \eta
)$.

We next outline the steps which result in the differential equation (2.6),
stated below.  Calculation of $(\partial /\partial s )R(s\vert \eta )$ brings
down under the integrals the quantity $i \lambda_a^I u_a (s)$, standing to the
left of its OE, which may be represented as $\lambda_a^I {\delta\over\delta
\alpha_a} (s)$ acting upon $\exp \left[ i\int\alpha \cdot \xi \right]$; then a
functional integration-by-parts moves this $\delta/\delta\alpha_a (s)$ to act
upon the last line of (2.3), which generates under the functional integrals the
net quantity $(-i)\int_{-\infty}^{+\infty} dt \lambda_a^I Q_{ab} (s,t) \beta_b
(t)$.  The procedure may now be reversed, representing $(- i)\beta_b (t)$ by
the operation $-{\delta/\delta v_b (t)}$ acting upon $\exp \left[ i \int v
\cdot \beta \right]$; and using another functional integration-by-parts to
convert this to the operation:

\begin{eqnarray}
& & {\delta\over\delta v_b (t)} \ \left[ \left( e^{i\int_{-\infty}^{+\infty}
\lambda^{\rr} \cdot v} \right)_+ \, e^{i\int v\cdot \eta} \right]\nonumber\\
& & = i \left[ \eta_b (t) + \left( e^{i\int_{t}^{\infty} \lambda^{\rr} \cdot v}
\right)_+ \lambda_b^{\rr} \, \left( e^{-i\int_{t}^{\infty} \lambda^{\rr} \cdot
v} \right)_- \right] \left( e^{i\int_{-\infty}^{+\infty} \lambda^{\rr} \cdot v}
\right)_+ \, e^{i\int v \cdot \eta }
\end{eqnarray}
written in terms of the anti-ordered quantity:

\[
\left( e^{-i\int_{t}^{\infty} \lambda^{\rr} \cdot v} \right)_-  =\left[  \left(
e^{i\int_{t}^{\infty} \lambda^{\rr} \cdot v} \right)_+\right]^{\dagger}  =
\left[ \left( e^{i\int_{t}^{\infty} \lambda^{\rr} \cdot v} \right)_+ \right]
^{-1}
\]

We introduce the notation:

\beq
\Lambda_b^{\rr} \left( t \vert iv \right) = \left( e^{i\int_{t}^{\infty}
\lambda^{\rr} \cdot v} \right) _+ \lambda_b^{\rr} \left( e^{-i\int_{t}^{\infty}
\lambda^{\rr} \cdot v} \right) _-
\eeq
and observe that (2.4) may be rewritten as:

\[
i \left[ \eta_b (t) + \Lambda_b^{\rr} \left( t \vert {\delta/\delta \eta}
\right) \right] \, \left( e^{i\int_{-\infty}^{+\infty} \lambda^{\rr} \cdot v}
\right)_+ \, e^{i\int v\cdot \eta}
\]
so that, finally, one obtains the differential equation:

\beq
{\partial R (s\vert\eta )\over \partial s} = i \int_{-\infty}^{+\infty} dt \,
\lambda_a^{I} Q_{ab} (s,t) \left[ \eta_b (t) + \Lambda_b^{\rr}\left(t \vert
{\delta\over\delta\eta}\right) \right] \cdot R (s\vert\eta )
\eeq

With the boundary condition $R (- \infty \vert \eta ) = 1$, easily seen as
appropriate from the definition of $R (s\vert\eta )$, the solution to (2.6) may
be written as an OE:

\beq
R (s,t) = \left(\exp \left[ i \int_{-\infty}^s \, ds^{\prime}
\int_{-\infty}^{+\infty} dt^{\prime} \lambda_a^I Q_{ab} \left( s^{\prime} ,
t^{\prime} \right)  \left[ \eta_b (t^{\prime} ) + \Lambda_b^{\rr} \left(
t^{\prime} \vert {\delta\over\delta\eta} \right)\right]\right]
\right)_{+s^{\prime}}
\eeq
with the ordering indicated for the $s^{\prime}$ variables only.  With
$s\rightarrow + \infty$ and $\eta \rightarrow 0$, we then have a representation
of (2.3) which apparently involves a single OE; however, it should be noted
that the second ordering will be found in the definition of $\Lambda_b^{\rr}$,
eq.(2.5), so that there do exist two sets of ``orderings", although they can
now be addressed separately.  In fact, the ``$t$-orderings" can be defined from
the integral solution to the differential equation satisfied by
$\Lambda_a^{\rr} \left( t \vert {\delta\over\delta\eta} \right) $; the latter
may immediately be obtained from its definition (2.5):

\[
{\partial\over\partial t} \Lambda_a^{\rr} \left( t \vert
{\delta\over\delta\eta} \right) = 2i \, f_{acd} {\delta\over\delta\eta_c (t)}
\, \Lambda_d^{\rr} \left( t \vert {\delta\over\delta\eta} \right)
\]
which, together with the boundary condition at $t=\infty$, generates:

\beq
\Lambda_a^{\rr} \left( t \vert {\delta\over\delta\eta} \right) =
\Lambda_a^{\rr} - 2i \, f_{acd} \int_t^{\infty} dt^{\prime} \,
{\delta\over\delta\eta_c (t^{\prime} )} \Lambda_d^{\rr} \left( t^{\prime} \vert
{\delta\over\delta\eta } \right)
\eeq
whose repeated iterations contain all the $t$-orderings of the problem, and
where the $f_{abc}$ are the structure constants of the SU(N) algebra.

Conventional eikonal models replace $Q_{a,b}$ by $\delta_{a,b} Q (s,t)$, and in
the absence of any other isospin/color vector, we may expect that the result
will generate the products $\lambda^I \cdot \lambda^{\rr} $.  The latter may
then be replaced by eigenvalues appropriate to the scattering problem; for
example, in the SU(2) isospin scattering of two nucleons, those eigenvalues are
given by I (I+1)/2-3/4, for singlet (I=0) or triplet (I=1) total isospin; for
SU(3), the situation is somewhat more complicated, as one tries to extract the
overall, contribution of the eikonal to the singlet scattering amplitude
\cite{C-W}.

While (2.7) is a formal solution of the problem, certain terms of its expansion
can be summed without difficulty.  To see this, consider the expansion of (2.7)
up to quadratic $Q$-dependence:

\begin{eqnarray}
& & R \mid_{s\rightarrow\infty}\simeq  \, 1 + i \int\!\int_{-\infty}^{+\infty}
ds^{\prime} dt^{\prime} \lambda_a^I Q_{ab} (s^{\prime} , t^{\prime} ) \left[
\eta_b (t^{\prime} ) + \Lambda_b^{\rr} \left( t^{\prime} \vert {\delta\over
\delta\eta} \right) \right]\, +\nonumber\\
& & + i^2 \int\!\int_{-\infty}^{+\infty} ds_1 dt_1 \lambda_a^I  Q_{a_{1}b_{1}}
\left( s_{1} , t_1 \right) \int_{-\infty}^{s_{1}} \, ds_2 \,
\int_{-\infty}^{+\infty} \, dt_2 \, \lambda_{a_{2}}^{\rr} \, Q_{a_{2} , b_{2}}
\left( s_2 , t_2 \right)\cdot  \\
& & \cdot \left[ \eta_{b_{1}} (t_1 ) + \Lambda_{b_{1}}^{\rr} \left( t_1 \vert
{\delta\over\delta\eta} \right)\right] \cdot \left[ \eta_{b_{2}} (t_2 ) +
\Lambda_{b_{2}}^{\rr} \left( t_2 \vert {\delta\over\delta\eta } \right)
\right]_{n\rightarrow 0} + \cdots \nonumber
\end{eqnarray}
With the definition of $\Lambda_b^{\rr} (t\vert {\delta\over\delta\eta} ) $, it
is clear that the only contribution of the linear $Q$-terms is:

\beq
i \int\!\int_{-\infty}^{+\infty} \, ds \, dt \, \lambda_a^I Q_{ab} (s,t)
\lambda_b^{\rr}
\eeq
while the ${\delta\over\delta\eta}$ -independent part of the quadratic
$Q$-terms of (2.9) yields:

\beq
i^2 \int\!\int_{-\infty}^{+\infty} ds_1 dt_1 \lambda_{a_{1}}^I Q_{a_{1} b_{1}}
Cs_1 , t_1 ) \lambda_{b_{1}}^{\rr} \int_{-\infty}^{s_{1}} \, ds_2
\int_{-\infty}^{+ \infty} \, dt_2 \lambda_{a_{2}}^I Q_{a_{2} b_{2}} \left( s_2
, t_2 \right) \lambda_{b_{2}}^{\rr}
\eeq
This structure, obtained from the first term, $\lambda_b^{\rr}$, in the
iterative expansion of $\Lambda_b^{\rr}$, eq. (2.8):

\beq
\Lambda_b^{\rr} \left( t \vert {\delta\over\delta\eta} \right) \simeq
\lambda_b^{\rr} - 2i \, f_{bcd} \lambda_d^{\rr} \int_t^{\infty} \, dt^{\prime}
{\delta\over \delta \eta_c (t^{\prime}} + \cdots
\eeq
will appear in every term of the complete expansion of $R$, and generates the
OE:

\beq
\left( \exp \left[ i \int\!\int_{-\infty}^{+\infty} ds \, dt \, \lambda_a^I
Q_{ab} (s,t) \lambda_b^{\rr} \right] \right)_{+(s)}
\eeq
If, as typical, $Q_{ab} = \delta_{ab} Q(s,t)$, all the $\lambda_a^I \cdot
\lambda_b^{\rr}$ terms in the expansion of (2.13) combine to form the products
$\lambda^I \cdot \lambda^{\rr}$, at which point the OE becomes an ordinary
exponential (oe):

\beq
\exp \left[ i \left( \lambda^I \cdot \lambda^{\rr} \right)
\int\!\int_{-\infty}^{+\infty} ds \, dt \, Q (s,t) \right]
\eeq
where the combination $\lambda^I \cdot \lambda^{\rr}$ may be replaced by its
appropriate eigenvalue.  The value of the integrals of (2.14) may be read off
from (1.3) and (1.4).
\par
It is instructive to continue with the example of (2.9) and calculate the first
commutator-term, as in (2.12), to this quadratic $Q$-dependence; it is:

\beq
2i f_{b_{1} b_{2} d} \int\!\int_{-\infty}^{+\infty} ds_1 dt_1 \lambda_{a_{1}}^I
Q_{a_{1}b_{1}}  \left( s_1 , t_1 \right) \int_{-\infty}^{s_{1}} d s_2
\int_{t_{1}}^{\infty} dt_2 \lambda_{a_{2}}^I Q_{a_{2}b_{2}} (s_2 , t_2 )
\lambda_d^{\rr}\, .
\eeq

If, again $Q_{a,b} = \delta_{a,b} Q(s,t)$, the antisymmetry of (2.15) under
$b_1 , b_2$ exchange is converted to a like antisymmetry under $a_1 , a_2$
exchange, so that the pair $\lambda_{a_{1}}^I \lambda_{a_{2}}^I$ may be
replaced by $if_{a_{1}a_{2}c} \lambda_c^I$.  One then finds the double
summation $\sum_{a_{1} a_{2}} f_{a_{1} a_{2} c }f_{a_{1} a_{2} d} = C_2
\delta_{cd}$, where $C_2 (N) = N$ denotes the value of the quadratic Casimir
invariant of the adjoint representation; and (2.15) becomes:

\beq
- 2 C_2 \left( \lambda^I \cdot \lambda^{\rr}\right)
\int\!\int_{-\infty}^{+\infty} ds \, dt \, Q (s,t) \int_{-\infty}^s \, ds_1
\int_t^{\infty} dt_1 Q (s_1 , t_1 )
\eeq

In a typical eikonal situation corresponding to NVM exchange,
$Q(s,t)= g^2 \hfill
(p_1 \cdot p_2 ) \Delta_c \left( z_1 - z_2 - sp_1 + t p_2 \right)$, and the
integrals of (2.16) may be evaluated to yield the leading $\ln (E/m )$
dependence:

\beq
i {C_2\over 2\pi} \left( {g^2\over\pi} \right)^2 \left( \lambda^I \cdot
\lambda^{\rr} \right) \ln ( E/m ) K_0^2 (\mu b )
\eeq
where $4E^2$ denotes the total CM (energy)$^2$ of the scattering quarks.  The
form of (2.17) is worth noting, for it contains the new feature of a $\ln
(E/m)$ dependence multiplying reasonable, impact-parameter dependence; as
explained in great detail in reference \cite{C-W}, it is the first appearance
of an effective Reggeization of the exchanged gluon, and it appears directly in
the eikonal function.

Before discussing how such contributions may be extracted and summed in this
functional context, it may be appropriate to note that there is at least one
kinematical context in which (2.13) is the exact result.  This is the special
case where $Q_{a,b} (s,t) = Q_{a,b} (s) \delta (s-t)$, when the functional
derivatives of (2.12) can never appear (due to a mis-ordering of subsequent,
proper-time variables).

Another example where differences may be expected from the usual eikonal forms
results from the appearance of a $Q_{a,b} = f_{abc} \xi_c Q(s,t)$, where
$\xi_c$ is a color vector in the flux-string model of reference \cite{Blois}.
Because this $Q$ is proportional to a delta function of the square of the $x_1
- x_2$ variables of (1.3), it produces an OE with only $s$ dependence, and the
kinematical forms which appear are quite different from the examples noted
above.

Other formulations of the solution to (2.6) are possible, such as the
representation of $R(s \vert \eta )$ by a Fourier functional transform, and the
subsequent conversion of (2.6) to a differential equation linear in parametric
and functional derivatives.  However, because of the non-commutation of the
$\lambda_a$, this route does not seem to lead to any real simplification.

\section{Contiguity}
\setcounter{equation}{0}
\indent

A representation for the general structure of all such terms may be obtained by
the following argument.  Return to the differential equation (2.6) for $R(s
\vert \eta )$ and make the ansatz: $R = R_0\, U_0$, where we shall assume in
all that follows that $Q	_{a,b} = \delta_{a,b} Q (s,t)$.  The quantity $R_0
(s)$ is defined by:

\beq
R_0 (s) \equiv \left( \exp \left[ i \int_{-\infty}^s \, ds^{\prime}
\int_{-\infty}^{+\infty} dt^{\prime} \lambda_a^I Q (s^{\prime} , t^{\prime} )
\left[ \eta_a (t^{\prime} ) + \lambda_a^{\rr} \right] \right]\right)_{+(s')}
\eeq
and substitution of (3.1) into (2.6) then yields:

\[
{\partial U_0\over \partial s} = i \int_{-\infty}^{+\infty} dt \, R_0^{-1} (s)
\lambda_a^I Q (s,t)\Delta \Lambda_a^{\rr} \left( t\vert
{\delta\over\delta\eta}\right)\, R_0 (s) \cdot U_0 (s),\,\, \Delta
\Lambda_a^{\rr} = \Lambda_a^{\rr} - \lambda_a^{\rr}
\]
with solution:

\beq
U_0 (s) = \left( \exp \left[ i \int_{-\infty}^s \, ds^{\prime}
\int_{-\infty}^{+\infty} \, dt^{\prime} \, R_0^{-1} (s^{\prime} ) \lambda_a^I
Q\left( s^{\prime} , t^{\prime} \right)\Delta \, \lambda_a^{\rr} \left(
t^{\prime} \vert {\delta\over\delta \eta} \right) R_0 (s^{\prime} )
\right]\right)_{+ (s')}
\eeq
from which we require the limits $s\rightarrow \infty , \eta \rightarrow 0$.
To quadratic order in $Q$, one finds that the expansion of $U_0$ generates
(2.16), as it must; but because of the $R_0$ factors inside the OE of (3.2),
higher-order terms will, at least in part, involve commutators of the
$\lambda$-dependence of $R_0$ with neighboring $\lambda^I , \lambda^{\rr}$
dependence of (3.2); those terms will be different from the simple
exponentiation of (2.16), but they will always be of higher perturbative order
then that of (2.16), and are not the leading terms of their own perturbative
order.  Note that the combination $\Delta\Lambda_a^{\rr}$ of (3.2) contains all
the multiple commutators, indicated in (2.12), whose functional derivatives act
upon the $\eta$-dependence of $R_0$.

To find that term in the eikonal of order $g^{2(n+1)}$ which is the leading
term of that order, let us now write:

\begin{eqnarray}
 \Delta \Lambda_a^{\rr} \left( t \vert {\delta\over \delta\eta} \right) &
\equiv & \sum_{n=1}^{\infty} \, \Delta_n \Lambda_a^{\rr} \left( t \vert
{\delta\over\delta\eta }\right) \, ,\nonumber\\
\Delta _n\Lambda_a^{\rr}  \left( t \vert {\delta\over\delta\eta}\right) & =
&\left( -2i \right)^n \, f_{ac_{1} d_{1}} \, f_{d_{1} c_{2} d_{2}} \cdots
f_{d_{n-1}, c_{n} d_{n}}\,\lambda_{d_{n}}^{\rr} \\
&&  \cdot \int_t^{\infty} \, dt_1 \int_{t_{1}}^{\infty} \, d t_2 \cdots
\int_{t_{n-1}}^{\infty} \, dt_n \, {\delta\over\delta\eta_{c_{1}} (t_1 )}
\cdots {\delta\over\delta \eta_{c_{n}} (tn)} \nonumber
\end{eqnarray}
and set $U_0 = R_1\, U_1$, where we define:

\[
 R_1 (s) = \left( \exp \left[ i \int_{-\infty}^s \, ds^{\prime}
\int_{-\infty}^{+\infty} \, dt^{\prime} \, R_0^{-1} (s^{\prime} ) \lambda_a^I Q
\left( s,^{\prime} t^{\prime} \right) \Delta_1 \Lambda_a^{\rr} \left(
t^{\prime} \vert {\delta\over\delta\eta} \right) R_0 (s^{\prime}
\right]\right)_{+(s')}
\]
Then, by again solving the appropriate differential equation, we find:

\begin{eqnarray}
& &  U_1 (s) = \left( \exp \left[ i \, \int_{-\infty}^s \, ds'
\int_{-\infty}^{+\infty} \, dt' \left[ R_0 (s') R_1 (s' )\right]^{-1} \,
\lambda_a^I Q (s,^{\prime} t^{\prime} )\right.\right.\nonumber\\
& & \cdot  \left.\left.\sum_{n=2}^{\infty} \Delta_n \Lambda_a^{\rr} \left(
t^{\prime} \vert {\delta\over \delta\eta}\right) \left[ R_0 (s') R_1 (s'
)\right]\right]\right)_{+(s')}\nonumber
\end{eqnarray}
Performing this operation sequentially, it is clear that the general structure
of the result may be written as:

\[
R \left( s\vert \eta \right) = R_0 (s) \cdot R_1 (s)\cdots R_n (s) \cdot U_n
(s) \equiv \left[ s_n \right] U_n (s)
\]
where:

\beq
R_n (s) = \left( \exp \left[ i \int_{-\infty}^s \, ds' \int_{-\infty}^{+\infty}
\, dt' [ s' ]_{n-1}^{-1} \, \lambda_a^I Q (s' , t' ) \cdot \Delta_n
\Lambda_a^{\rr} \left( t' \vert {\delta\over\delta \eta} \right) [s' ]_{n-1}
\right] \right)_{+(s')}
\eeq
and:

\beq
U_n (s) = \left( \exp \left[ i \int_{-\infty}^s \, ds' \int_{-\infty}^{+\infty}
dt' \, [s']_n^{-1} \lambda_a^I Q (s' , t' ) \cdot \sum_{\ell = n + 1}^{\infty}
\, \Delta_{\ell} \Lambda_a^{\rr} \left( t' \vert {\delta\over\delta \eta
}\right) [ s' ]_n \right] \right)_{+ (s')}
\eeq
Because each functional derivative ${\delta\over\delta\eta}$ will generate a
term (when operating on $R_0 (s)$) proportional to $Q \sim g^2 \Delta_c$, the
log of $R_n$ contains all powers of $g^{2m}$, with $m\geq n+1$.  The lowest
order term, with $m=n+1$, will contain the largest power of $\ln^n (E/m )$,
while higher-order terms constructed from the same $R_n$ will have no
higher-order log; rather, the terms containing $\ln^m (E/m ) , m> n+1$, will
come from the corresponding, lowest-order terms of $R_m$.
\par
In order to define ``contiguity", imagine that $R_n$ is expanded in powers of
$g^2$, by expanding its OE:

\begin{eqnarray}
& & R_n \mid_{s\rightarrow\infty} \simeq 1 + i \int\!\int_{-\infty}^{+\infty}
\, dsdt\,\, [s]_{n-1}^{-1} \lambda_a^I Q_{ab} (s,t) \Delta_n \Lambda_a^{\rr}
\left( t \vert {\delta\over\delta\eta} \right) [s]_{n-1} \nonumber\\
& & + i^2 \int\!\int_{-\infty}^{+\infty} dsdt\,\, [s]_{n-1}^{-1} \lambda_a^I Q
(s,t) \Delta_n\Lambda_a^{\rr} \left( t \vert {\delta\over\delta\eta}\right)
[s]_{n-1} \\
& & \cdot \int_{-\infty}^s \, ds_1 \int_{-\infty}^{+\infty} \, dt_1 [ s_1
]_{n-1}^{-1} \, \lambda_{a_{1}}^IQ (s_1 , t_1 ) \Delta_n \Lambda_{a_{1}}^{\rr}
\left( t_1 \vert {\delta\over\delta\eta} \right) [s_1 ]_{n-1}
\,+\cdots\nonumber
\end{eqnarray}
where $[s]_{n-1} = R_0 (s) \cdots R_{n-1} (s)$.  ``Contiguity" suggests that
the leading dependence of $\ln (R_n )$ will be obtained if each $\Delta_n
\Lambda_{a_{j}}^{\rr} \left( t_j \vert {\delta\over\delta\eta}\right)$ operates
directly upon the $[s_j ]_{n-1}$ factor contiguous to it, that is, immediately
to its right.  This can be seen in the simplest, non-trivial terms of order
$g^4$ and $g^6$, and, we subsequently argue, is true for all terms; however,
what is clear from this definition is that   terms contributing to each order
of the contiguity operation can be summed and calculated directly from the OE
form of $R_n$, writing:

\beq
R_n \mid_{s\rightarrow \infty} = \left( \exp \left[ i
\int\!\int_{-\infty}^{+\infty} \, dsdt\,\, [s]_{n-1}^{-1} \, \lambda_a^I Q
(s,t) \underbrace{\Delta_n \Lambda_a^{\rr} \left( t \vert
{\delta\over\delta\eta}\right) [s}]_{n-1} \right]\right)_{+(s)}
\eeq
where the factor-pairing notation is meant to express the subset of terms
extracted by contiguity.

The entire $g^{2n}$ dependence of the eikonal, that is, of $\ln (R)$, can be
obtained by considering the following sequence of ascending powers of $g^2$, in
the limit of $s\rightarrow \infty , \eta \rightarrow 0$:

All $g^2$ dependence is given by $R_0 , \ln (R_0 ) = i \left( \lambda^I \cdot
\lambda^{\rr} \right) \int\!\int_{-\infty}^{+\infty} \, dsdt Q (s,t)$.
\par
All $(g^2)^2$ dependence is given by the contiguity calculation of $R_1$, which
generates our previous result, $\ln (R_1 ) = - 2C_2 \left( \lambda^I
\cdot\lambda^{\rr} \right) \int\!\int_{-\infty}^{+\infty} \, dsdt Q (s,t)
\int_{-\infty}^s \, ds_1 \int_t^{\infty} \, dt_1 Q (s_1 , t_1 ) $.
\par
All $(g^2)^3$ dependence is given by the contiguity calculation of $R_2$, and
by the $g^2$ expansion of the $[s]_0^{-1}$ and $[s]_0$ factors of $R_1$.
\par All $(g^2 )^4$ dependence is given by the contiguity calculation of $R_3$,
by the $g^2$ expansion of the $[s]_1^{-1}$ and $[s]_1$ factors of $R_2$, and by
the $g^4$ expansion of the $[s]_0^{-1}$ and $[s]_0$ factors of $R_0$; etc.

In this way, one constructs the complete $g^{2(n+1)}$ dependence of $\ln (R) =
\ln (R_0 \cdots R_n)$.  Those exponential, eikonal terms obtained directly from
contiguity will contain one or more terms proportional to a single factor of
$\lambda^I \cdot \lambda^{\rr}$, while the $g^{2p}$ expansions of the $[s_j]$
and $[s_i]^{-1}$ appear to generate more complicated group factors, similar to
those found in the perturbative calculations of the amplitude \cite{C-W}. We
argue in the next Section that the leading $\ln (E/m)$ dependence to the
eikonal of order $g^{2(n+1)}$ comes only from the contiguity calculation of
$R_n$, when the functional differentiation is performed only on the $R_0(s)$
factor of $[s]_n$.  Using simple functional techniques, the sum of these
leading contributions over all orders $ n$ is constructed for the eikonal of
SU(2).

\section{Leading Logs}
\setcounter{equation}{0}
\indent

We here give a qualtitative discussion of the leading $\ln (E/m)$ dependence of
this class of non-abelian eikonals (where, we again remind the reader,
interacting gluons are not included).  For this, consider first those terms of
order $g^{2(n+1)}$ in the expression for $\ln (R_n )$ arising from the
contiguity operation of $\Delta_n \Lambda_a^{\rr}$ upon the factor $[s]_{n-1}$
standing to its immediate right, as in (3.7).  In particular, the leading terms
of that order will come from the $\Delta_n \Lambda_a^{\rr}$ operation upon the
$R_0 (s)$ functional in $[s]_{n-1}$ (rather than the same-$g^2$-order
contributiuon to the eikonal from $\ln (R_{n-1})$, with $\Delta_{n_1}
\Lambda_a^{\rr}$ acting upon $R_1 (s)$ in $[s]_{n-2}$, etc).

For clarity, we carry the discussion through for $n=2$, and then generalize to
arbitrary $n$; for the moment, we suppress the $f_{abc}$ factors arising in the
$t$-dependent iterations of $\Lambda_a^{\rr} \left( t\vert
{\delta\over\delta\eta}\right)$, but we explicitely write the $s$-dependent
permutations that are generated by the functional differentiation of $\Delta_2
\Lambda_a^{\rr} \left( t\vert {\delta\over\delta\eta}\right)$ upon $R_0(s)$,
which are proportional to:

\beq
\int_t^{\infty} \, dt_1 {\delta\over\delta\eta_{c_{1}} (t_{1}
)}\int_{t_{1}}^{\infty} dt_2 {\delta\over \delta\eta_{c_{2}} (t_2)} \left( \exp
\left[ i \int_{-\infty}^s \, ds' \int_{-\infty}^{+\infty} \, dt' Q (s' , t' )
\lambda_a^I \eta_a (t') \right]\right)_{+(s')} \mid_{\eta\rightarrow 0}
\eeq
We have neglected in this $R_0(s)$ its exponential $i \left( \lambda^I \cdot
\lambda^{\rr} \right) \int_{-\infty}^s \, ds' \, \int_{-\infty}^{+\infty} \,
dt' Q (s',t')$ dependence because, as explained below, it can only contribute
to orders $g^{2p} , p > n+1$, and carries no additional $\ln (E/M)$ factors.
Suppressing the superscript $I$ for each $\lambda_c^I$, the functional
operations of (4.1) yield:

\beq
i^2 \int_t^{\infty} dt_1 \int_{t_{1}}^{\infty} \, dt_2 \int_{-\infty}^s \, ds_1
\int_{\infty}^s \, ds_2 Q (s_1 , t_1 ) Q (s_2 ,t_2 ) \left[ \lambda_{c_{1}}
\lambda_{c_{2}} \theta (s_1 - s_2 ) +\lambda_{c_{2}} \lambda_{c_{1}} \theta
(s_2 - s_1 )\right]
\eeq
and suggest the obvious generalization for $n> 2$ as:

\beq
i^n \int_t^{\infty} \, dt_1 \cdots \int_{t_{n-1}}^{\infty} \, dt_n \sum_{perm}
\int_{-\infty}^s \, ds_1 \cdots \int_{-\infty}^{s_{n-1}} \, ds_n \cdot
\lambda_{c_{1}} \cdots \lambda_{c_{n}} \, Q (s_1, t_1 ) \cdots Q (s_n , t_n )
\eeq
in which the $n$  $c_i$ indices are permuted, with a corresponding permutation
of the $s_i$, in $n!$ different ways.

For our estimates of the $\ln (E/m)$ depenence, we use the standard Fourier
propagator representation, $\Delta_c (x) = (2\pi )^{-4} \int d^4 k (k^2 + \mu^2
- i\epsilon)^{-1} e^{ik\cdot x}$, and (improperly) take the kinematic limits
for each (mass-shell) quark:  $E - p = 0$, rather than the more accurate $E - p
\simeq m^2/2E$.  Any integral that we find containing an $UV$ log divergence is
really proportional to a corresponding factor of $\ln (E/m )$, which dependence
appears when proper (but much more complicated) kinematics are used.

For $n=2$, let us examine both permutations, and include the $i\lambda_a^I
\int\!\int ds \, dt \, Q (s,t)$ factor of (3.7), whose $[s]_1^{-1}$ has been
replaced by unity (because it can only contribute to higher orders with no
corresponding increase in the number of $\ln (E/m)$ factors).  Each factor of
$Q$ carries with it $p_1 \cdot p_2 \sim E^2$ dependence, which is removed by
the explicit $E$-factors associated with the $s$- and $t$-integrations, in
standard eikonal fashion; and we suppress all such cancelling $E$-dependence.
With $Q (s,t) = g^2 \left( p_1\cdot p_2 \right) \Delta_c \left( z - sp_1 + tp_2
\right), $ where $z = z_1 - z_2 = (b, z_3 , z_0 )$ is the difference of
configuration coordinates of the scattering quarks, the first of the two
permutations of (4.1) will lead to:

\begin{eqnarray}
& & \int d^4\bar{k}^{(+)} \, e^{i\bar{k} z} \delta \left( \bar{k}^{(+)} \right)
\delta \left( \bar{k}^{(-)} \right) \int d^2 k_1 \int d^2 k_2 \int dk_1^{(+ )}
\int dk_1^{(-)} \int dk_2^{(+  )} \int dk_2^{(-)} \nonumber\\
& &\left[ \omega^2 \left( \bar{k} - k_1 - k_2 \right) + \left( k_1^{(+)} +
k_2^{(+)} \right)\left( k_1^{(-)} + k_2^{(-)} \right) - i\epsilon \right]^{-1}
\, \left[ w_1^2 + k_1^{(+ )} k_1 ^{(-)} - i\epsilon\right]^{-1}\nonumber\\
& & \left[ \omega_2^2 + k_2^{(+)} k_2^{(-)} - i\epsilon\right]^{-1} \left(
k_2^{(-)} + i\epsilon \right)^{-1} \left( k_1^{(-)} + k_2^{(-)} +
i\epsilon\right)^{-1} \, \left( k_2^{(+)} - i\epsilon \right)^{-1} \, \left(
k_1^{(+ )} + k_2^{(+ )} - i\epsilon\right)^{-1} \nonumber
\end{eqnarray}
where $\bar{k} = k + k_1 + k_2, k^{(\pm )} = k_3 \pm k_0 , \omega^2 = \mu^2 +
k_{\bot}^2$, and $\omega^2 \left( \bar{k} - k_1-k_2 \right)^2 = \mu^2 + \left(
\bar{k} - k_1 - h_2 \right)_{\bot}^2$,
with $\bot$  components referring to the transverse 1,2 directions
(the impact parameter vector $b$) in the CM of the scattering
quarks; all momentum integrals run from $-\infty$ to $+\infty$.  The $  ±i
\epsilon$
factors are important, and - aside from the -$i\epsilon$    of the standard
Feynman propagator denominators - arise upon calculating
$ \int_{-\infty}^s ds_1 \int_{-\infty}^{s_{1}} ds_2$  and $\int_t^{\infty} dt_1
\int_{t_{1}}^{\infty} dt_2$, when one insists upon the proper
definition of the integrand at the $\pm\infty$ limits of integration.  The
second permutation of (4.1) leads to the same form with the
interchange of $k_1^{(-)}$ and $k_2^{(-)}$, and, it will become clear
immediately, to the same leading-log dependence.

It is best to begin by performing the $\int d k_{1,2}^{(-)}$
integrations, which by simple contour evaluation require $k_1^{(+)} >
0$ and $k_1^{(+)} + k_2^{(+)} > 0$, and generate:

\begin{eqnarray}
& & \left( - 2\pi i\right)^2 \int {d^2k\over\omega^2} \, e^{ib\cdot k_{\bot}}
\int {d^2 k_1\over \omega_1^2} \, e^{i b\cdot k_{1}} \, \int {d^2 k_2\over
\omega_1^2} \, e^{ib\cdot k_{2}} \nonumber\\
& & \cdot \int_{\epsilon}^K \, {dk_2^{(+)}\over k_2^{(+ )} } \,
\int_{\epsilon}^K \, {dk_1^{(+ )}\over k_2^{(+ )} + k_1^{(+ )}}  \left\{ 1 + \,
{k_2^{(+ )} \omega_1^2\over k_2^{(+ )} \omega_1^2 + k_1^{(+ )} \omega_2^2
}\right\}
\end{eqnarray}
where we have inserted upper $(K) $ and lower $(\epsilon  )$ cut offs for
the $k_{1,2}^{(+)}$ integrations, and have replaced the transverse
$\bar{k}_{\bot}$
variable by
$(k + k_1 + k_2)_{\bot}$   .   Each of the three factors $\int d^2 k\omega^{-2}
\, e^{ik\cdot b}$
generates a term $(2\pi)  K_0 (\mu b)$, and the``1" of the curly bracket
of (4.4) produces a ``nested" contribution for the $k_{1,2}^{(+)}$
integrals of amount $(1/2) \ln^2(K/\epsilon )\rightarrow (1/2) \ln^2(E^2/m^2)$,
when the replacement $E - p_3 \simeq    m^2/2E$ is used.  In contrast, the
second term of the curly bracket of (4.4) generates a
contribution proportional to $ln(E/m)$, and can be dropped as
sub-leading.  Quite generally, a ``nesting" of the $k_i^{(+)}$ momenta
follows directly from the ordered $t$-limits of the iterates of
 $\Delta_n \Lambda_a^{\rr}$, while the sum over all permutations of the
$\lambda_{c_{1}} \cdots \lambda_{c_{n}}$ follows
from the ordered $s$-limits of the terms obtained upon
functional differentiation of $R_0(s)$ by $\Delta_n \Lambda^{\rr}$.
The leading-log result for each $\lambda_{c_{1}} \cdots\lambda_{c_{n}}$
     permutation is proportional to $(1/n!) \ln^n(E^2/m^2)$.

One can easily see that any expansion of the $i
\left( \lambda^I \cdot \lambda^{\rr} \right) \int_{-\infty}^{+\infty} ds
\int_{-\infty}^{+\infty} dt Q (s,t)$ portion of the exponent of $R_0 (s)$,  in
conjunction with the above forms, must always produce sub-leading
dependence, because at least one of the nested $k^{(+) }$
denominators needed for leading-log behavior will be missing.
Further, one can also see the reason for the importance of the
contiguity prescription, for - when the OEs defining each $R_n$ are
each expanded in powers of $g^2$ - all the non-contiguous $\Delta_{\ell}
\Lambda^{\rr} \left( t\vert {\delta\over\delta\eta}\right)$
operations will display ``improper", or out-of-sequence limits
for the $s$-integrals, which will generate a similar sort
of sub-leading behavior.  For this standard choice of propagator,
contiguity generates a first sub-division of terms containing
the desired, leading-log dependence; and the latter are then
isolated by the retention of only $R_0(s)$ in each factor of $[s]_{n-1}$,
and the neglect of every $[s]_{n-1}^{-1}$, in each $R_n(s)$.

Perturbative eikonal analyses quite similar to the above
have appeared long ago, in connection with multiperipheral
processes of scalar ``tower" exchange.  There also one expects
$k_i^{(-)}\sim   0$ and large, nested, $k_i^{(+)}$ momenta.  What is different
here (aside from trivial, complex, multiplicative factors) is that
one must also include the sums over all $\lambda_{c_{1}} \cdots
\lambda_{c_{1}}$              permutations, and the general form of such a sum
is not clear for SU(N).

For SU(2), however, this computation can be carried through,
and we now sketch that calculation.  Its essence is to replace the
leading-log dependence by another method of extraction which
does not arise from the nested $k_i^{(+)}$ integrations, but yields,
term-for-term and order-by-order, the same results.  This
method is defined by retaining the same $f_{abc}$ factors obtained
from the $\Delta_n \Lambda_a^{\rr}$ iterations, and multiplying those that
contribute to order $n$ by the terms:

\beq
{i^n\over n!} \sum_{perms} \, \int_{-\infty}^s \, ds_1 \cdots
\int_{-\infty}^{s_{n-1}} \, ds_n \int_t^{\infty} \, dt_1 \cdots
\int_t^{\infty} \, dt_n \, Q\, (s_1 , t_1 )\cdots Q (s_n , t_n )
\lambda_{c_{1}} \cdots \lambda_{c_{n}} \hfill
\eeq
and then summing over all $n$.  It is easy to see that the leading-
log contributions of (4.5) are identical to those of (4.4); the
only difference is that the $\int_{\epsilon}^K dk^{(+)}/k^{(+)}$
contributions of (4.5) are not nested, and that the $(n!)^{-1} $ which follows
from (4.4)
because of nesting is, in (4.5), inserted by hand.  This
replacement can be made for arbitrary SU(N); but the next step,
summing over all permutations of the $s$-ordering, seems to be
straightforward only for SU(2).

Because the $t$-integrals of (4.5) are not ordered, we
introduce the symbol $A_c(t) = \int_t^{\infty} dt'  {\delta\over\delta\eta_c
(t')}$ , and $\sum_a^{\rr}[A]$ as the sum
of all functional operations, which when performed on $R_0(s)$,
generate the correct sequence of $\epsilon_{abc}$ coefficients
multiplying (4.5).   $\Delta \sum^{\rr} [A]$ corresponds to the set of all the
iterations of the SU(2) version of $\Delta\Lambda^{\rr} \left(
t\vert{\delta\over \delta\eta}\right)$, where a factor of $(n!)^{-1}$ is
inserted for each $n$th order, and the operators $ A_{c_{1}} \cdots A_{c_{n}} $
 replace the $t$-ordered $\int_t^{\infty} dt_1\int_{t_{1}}^{\infty}dt_2 \cdots
\int_{t_{n-1}}^{\infty} dt_n {\delta\over\delta\eta_{c_{1}} (t_1 )} \cdots
{\delta\over\delta\eta_{c_{n}} (t_n)}$  of the expansion
of (2.8). We work directly with the contiguity approximation to
$U_0$ (rather than to the $R_n$ separately), in which $R_0^{-1}$ is replaced by
unity, and the leading-log simplification of $R_0$ is used, as in (4.1);
everywhere, the $f_{abc}\rightarrow \epsilon_{abc}$, and $\lambda_c\rightarrow
\sigma_c$.  One may
now examine the first four terms of this expansion, and it then
becomes clear, by inspection, that the full sum over all such
$A$-dependence may be written as:

\beq
\Delta\sum_a [A] = \left[ A^2 \delta_{ab} - A_a A_b\right] \sigma_b \cdot
{1\over A^2} \left\{ \cosh (A) - 1 \right\} - i \epsilon_{acd} \sigma_d \cdot
A_c {\sinh (A)\over A}
\eeq
where $A^2 = \sum_c A_c^2$, and $A = [A^2]^{1/2}$.  To obtain (4.6), one
repeatedly uses the SU(2) property $\sum_c \epsilon_{abc} \epsilon_{cde} =
\delta_{ad} \delta_{be} - \delta_{ae} \delta_{bd}$.

 The functional expression of contiguity, of $\Delta \sum_a^{\rr} [A]$
    operating on $R_0(s)$, can be performed by first introducing the
representations:

\beq
A_a \, {\sinh (A)\over A} = {1\over 2\pi} \int \, d^3 u \, \delta \left(
\vec{u}^2 - 1 \right) \, {\partial\over\partial u_a} \, e^{\vec{u}
\cdot\vec{A}}
\eeq
and
\begin{eqnarray}
& & \left[ A^2 \delta_{ab} - A_a A_b \right] \, {1\over A^2} \left\{ \cosh (A)
- 1 \right\}\nonumber\\
& & = {1\over 2\pi} \int_0^1 \, {d\lambda\over\lambda} \int d^3 u \, \delta
\left( \vec{u}^2 - 1\right) \left[ \delta_{ab} \left(
{\partial\over\partial\vec{u}}\right)^2 - {\partial\over\partial u_a}
{\partial\over \partial u_b }\right] e^{\lambda\vec{u} \cdot\vec{A}}
\end{eqnarray}
where $(\lambda   ,{\bf u})$ are dummy integration variables.  The quantity
$e^{\lambda\vec{u} \cdot \vec{A}} R_0 (s)\vert_{\eta\rightarrow 0}$ is then the
OE:

\beq
\left( \exp \left[ i \int_{-\infty}^s \, ds' \int_t^{\infty} dt' \,Q (s' , t' )
\lambda \left( \vec{\sigma}^I \cdot\vec{u} \right)\right]\right)_{+ (s') }
\eeq
and may be replaced by the oe:

\beq
\exp \left[ i \lambda \left( \vec{\sigma}^I \cdot\vec{u}\right) \, K (s,t)
\right]
\eeq
where $K(s,t) = \int_{-\infty}^s ds'\int_t^{+\infty} dt' Q(s',t')$.  In effect,
the lack of $A$-ordering
for these leading-log terms has transformed their operation
upon $R_0(s)$ into an ordinary exponential with weightings to be
determined by the integrations of (4.7) and (4.8).  These last
steps are now easily performed, by the replacement of (4.10)
by $\cos \left( \lambda u K \right) + i\left( \vec{\sigma}^I \cdot
\vec{u}\right){\displaystyle{\sin (\lambda uK )\over u}}$,
and its substitution into (4.7) and (4.8), whose evaluations yield:

\beq
A_a {\sinh (A)\over A} \cdot R_0 (s) \vert_{\eta\rightarrow 0} = {i\over 3} \,
\sigma_a^I \left( K \cos K + 2 \sin K \right)
\eeq
and:

\beq
\left[ A^2 \delta_{ab} - A_a A_b \right] {1\over A^2} \left\{ \cosh (A) -
1\right\} R_0 (s) \vert_{\eta\rightarrow 0} = {4\over 3} \delta_{ab} \left[
\cos (K) - 1 -  {K\over 2} \sin (K) \right]
\eeq
 From (4.6) and (4.10), (4.11) and (4.12), one obtains:

\begin{eqnarray}
\Delta \sum_a^{\rr} [A] R_0 (s) \vert_{\eta\rightarrow 0} & = & {1\over 3} \,
\epsilon_{acd} \sigma_d^{\rr} \sigma_c^I \, \left[ K \cos K + 2 \sin K\right]
\nonumber\\
& + & {4\over 3 }\sigma_a^{\rr} \left[ \cos K - 1 - {K\over 2} \sin K\right]
\end{eqnarray}
Multiplying (4.13) on the left by $\sigma_a^I$  , antisymmetrizing where
appropriate (together with the Casimir relation $\sum_{ac} \epsilon_{acd}
\epsilon_{ace} = 2 \delta_{de} )$,                         ),
and including the $R_0$ contribution of the product $R=R_0\,U_0$, one
finds the eikonal given by:

\begin{eqnarray}
\chi & = & \left( \sigma^I \cdot \sigma^{\rr} \right)
\int\!\int_{-\infty}^{+\infty} dsdt\, Q\, (s,t) \left\{ 1 - {4\over 3} \left[ 1
- \cos K + {K\over 2} \sin K \right]\right. \nonumber\\
& + &\left. {2\over 3} i \left[ K \cos K + 2 \sin K \right]\right\}
\end{eqnarray}
Finally, if one imagines expanding (4.14) in powers of $K(s,t)$,
and combines each $K^n(s,t)$ with the remaining integrand of $U_0$,
one may use the easily-verified property, correct for the
leading-log dependence of each order:

\beq
\int\!\int_{-\infty}^{+\infty} dsdt \, Q\, (s,t) K^n (s,t) \simeq \left[ - i
{g^2\over\pi^2} \, \ln (E/m) K_0 (\mu b ) \right]^n \equiv [ - i L]^n
\eeq
so that, upon resumming these terms into the equivalent of
(4.14), in effect the quantity $K(s,t)$ may be replaced by -$iL$ of
(4.15), yielding:

\beq
\chi = - {g^2\over 2\pi} \left( \sigma^I \cdot \sigma^{\rr}\right) K_0 (\mu b )
\left\{ 1 - {4\over 3} \left[ 1 - e^L \right] + {2\over 3} L \, e^L\right\}
\eeq
as the complete eikonal in leading-log approximation for the
SU(2) problem (e.g., of nucleon-nucleon scattering by the
exchange of neutral and charged vector mesons, with
conserved isospin).

Perhaps the most obvious feature of (4.16) is its
proportionality to $\sigma^I \cdot \sigma^{\rr}$, which quantity takes on
isoscalar or
isovector eigenvalues depending on the nature of the initial
scattering states.  A second interesting property is that, by
expressing the $\exp[L]$ factors of (4.16) in terms of:
\[
e^L = \left( s/m^2 \right)^{{g^2\over 2\pi^{2}} K_{0} (\mu b)}
\]
one finds an ``effective Reggeization" of the eikonal, where s
here again denotes total CM (energy)$^2$.  For $\mu\not= 0$, there is little
contribution to the scattering amplitude for small $b$; and hence
if $K_0(\mu b)$ is approximated as $\sim\exp[-\mu b]$, one obtains forms
similar to those found in the Regge-eikonal approximation of
multiperipheral scattering, except that this eikonal is real.  In
fact, the amplitude, constructed in the generic form (and
suppressing all inessential factors):

\beq
T \sim i s \int_0^{\infty} bdb \, J_0 (qb) \cdot \left[ 1 - e^{i\chi
(s,b)}\right]
\eeq
exhibits a variant of a ``hard disc" scattering solution, in that
there are two regions of impact parameter, $b\mathop <\limits_{>} b_0$, which
produce different contributions to the amplitude.  This can be
seen by defining $b_0$ as that impact parameter where $L(b_0) = 1,
 b_0 = \mu^{-1} \ln \left[(g^2/2\pi^2) Y \right] > \mu^{-1},
Y = 2 \ln(E/m)$, and writing the contributions to the amplitude of
(4.17) in terms of integrations over these two regions of $b$.
Since $L(b) = \exp[\mu (b_0 -b)]$, and we assume that $Y$ is large, when
$b < b_0, L$ is large, as is the eikonal of (4.16), and the only
significant contribution to the amplitude comes from the "1" of
the first term of (4.17).  When $b > b_0$, $L$ is small, and the only
significant contribution to the eikonal comes from the ``1" of
the bracket of (4.16), which we shall call $\chi_0$; this is the
contribution coming from the original $R_0 $  term of (2.14). This
argument leads to the representation of the amplitude of
(4.17) as the sum of two parts:

\beq
T \sim is \int_0^{b_{0}} \, bdb \, J_0 (q	b) + is \int_{b_{0}}^{\infty} bdb \,
J_0 (qb) \left[ 1 - e^{i\chi_{0}} \right]
\eeq
or as:

\beq
T\sim is \int_0^{b_{0}} bdb \, J_0 (qb) e^{i\chi_{0}} + is \int_0^{\infty}
bdb\, J_0 (qb) \left[ 1 - e^{i\chi_{0}}\right]
\eeq
in which the amplitude is characterized by by its simplest
eikonal approximation, $\chi_0$, and by the range parameter
$b_0(E/m)$ which defines that impact parameter beyond which
leading-log corrections force the eikonal to become extremely
large and oscillatory, thereby removing its contribution from the
amplitude.

Could the same mechanism be operative for the
general case of SU(N)?  Even though we cannot perform the
closed sum over all orders of leading-log contributions for $N >
2$, one can anticipate that for a similar $b_0(E/m)$ the eikonal
becomes very large, contributing a rapidly oscillating and
negligible contribution to the amplitude, which may be written
in the form of (4.18) or (4.19), with the  $\sigma^I \cdot\sigma^{\rr}$
invariant of $\chi_0$
replaced by $\lambda^I\cdot \lambda^{\rr}$. We think it a reasonable conjecture
that this
simple form is the actual result of the complete SU(N)
calculation.  Of course, this point is somewhat academic, since
when energies are large enough to take leading-logs seriously,
other processes which have here been neglected (e.g.,
multiperipheral production) are going to appear. Nevertheless,
it is of some theoretical interest to examine an amplitude
constructed from the eikonal of (4.16), under the assumption
that $\ln(E/m) >>1$; and it will be most interesting to see if similar
structures and simplifying approximations are going to appear
in the study of other eikonal processes which reflect the
growth of inelastic particle production with increasing
energies.

\section{Other Processes}
\setcounter{equation}{0}
\indent

An important variation of the non-abelian eikonal
scattering problem is found when self-energy processes (as in
radiative corrections to other QCD $n$-point functions) are
attempted.  Here, one may make use of the new, exact and
approximate representations for the needed Green's functions
of reference \cite{HMF-YG 1} in which dependence on the source fields,
$A_{\mu}$  and $F_{\mu \nu}$ is that of an OE of linear form; for the simplest
example, we omit the  $F_{\mu \nu}$ terms, and work in a quenched
approximation, so that the sum of all radiative corrections to
the fermion propagator will require evaluation of the quantity:

\begin{eqnarray}
R\left( s\vert \xi \right) &=& N' \int d[u] \int d[\alpha ] \, e^{i\int
\alpha\cdot u} \left( e^{i\int_{-\infty}^{s} \lambda \cdot u }\right)_+
\nonumber\\
& & e^{{i\over 2} \int\!\int \alpha_{a} Q_{ab} \alpha_{b}} \cdot e^{i\int
u\cdot\xi }
\end{eqnarray}
in the limit of $ s\rightarrow \infty$ and $\xi_a(s') \rightarrow 0$. Here,
$Q_{a,b}(s,t)$ is considerably more complicated than the corresponding function
of an eikonal scattering amplitude (although the resemblance
becomes closer if an improper, no-recoil approximation is
adopted), but must satisfy $Q_{ab}(s,t) = Q_{ba}(t,s)$.

Using techniques modeled after those sketched above, it
is easy to see that a representation of (5.1) is given by the
formal OE:

\beq
R \left( s \vert \xi\right) = \left( \exp \left[ i \int_{-\infty}^s ds'
\int_{-\infty}^{+\infty} dt' \lambda_a Q_{ab} (s' , t' )\left[ \xi_b (t' ) +
\theta (s' - t' )\Lambda_b \left( s' , t' \vert {\delta\over\delta\xi } \right)
\right]\right]\right)_{+(s')}
\eeq
where $ \Lambda_b (s,t \vert iu ) = \left( e^{i\int_{t}^{s} \lambda\cdot u}
\right)_+ \, \lambda_b \left( e^{-i \int_{t}^{s} \lambda\cdot u} \right)_- $.
The same, formal expansion corresponding to (3.4) and (3.5)
may be defined, except that $R_0 $ is now multiplied by the
exponential factor $\exp[ {i\over 2} \int \xi Q \xi ]$, which has the effect of
inserting polynomial $\xi$-dependence into all the exponents of subsequent
$R_n$, and the power-counting arguments given above must be
appropriately modified.

Perhaps the most interesting generalization of the forms
of Section III should appear in eikonal quark-scattering models
when  gluon-gluon interactions (e.g., the "tower graphs" and
their generalizations) are taken into account.  Before a
functional treatment can be attempted, even in the relatively
simple models described in the last chapters of references \cite{C-W}
and \cite{HMF B1}, it is necessary to have a decent representation - as a
functional of an equivalent gluon source used to represent
internal, ``$s$-channel" gluon exchanges - for the Green's
function corresponding to the $t$-channel gluons exchanged
between quarks. For the eikonal situation where different
spin-one bosonic fields are used to describe distinct $t$- and $s$-
channel exchanges, respectively, such a representation now
exists \cite{HMF-YG 2}, and can be written down without undue
complications; for the single gluonic field of real QCD, the
situation is similar but not as straightforward.

If these calculations can be carried through for the tower
graphs (corresponding to two-gluon, $t$-channel exchange
between scattering quarks) in a functional context, using
contiguity as  appropriate, there should then be an immediate
functional generalization which includes multiple, $t$-channel
gluon exchanges.  Such estimates of the QCD eikonal would be
most relevant to high-energy particle scattering experiments.

\section{Summary and Acknowledgements}
\indent

In this paper we have shown how the formidable,
non-abelian eikonal combination (1.1) may be written as the OE $R(s|\eta
) $ in the limit as $s\rightarrow \infty$, and $\eta\rightarrow 0$; and have,
by contiguity, isolated a sub-set of terms which exponentiate and contribute
directly to the eikonal function, and which contain appropriate
$\ln(E/m)$ dependence associated with the leading-log behavior
of every perturbative order. For SU(2), these terms may be
summed to all orders, generating an eikonal dependent on the
total isospin of the scattering channel, which displays a form
of Reggeization peculiar to this set of graphs summed.

Contiguity may also be phrased in terms of the original
ansatz,  $ R(s|\eta   ) = R_0\,U_0$, by replacing the exact $U_0$ of (3.2) by
its contiguity approximation, as used for the SU(2) calculation.
However, at least for the specifically perturbative estimates of
$\ln(U)$, it appears to be simpler to adopt contiguity in the
context of the $R_n$. As explained in Section III, contiguity
together with the elimination of obviously sub-leading terms,
provides a straightforward method for the estimation of the
eikonal's leading-log terms in every perturbative order.  We
have found an elementary method for summing all such terms
in SU(2), and conjecture the form of a simplified eikonal
amplitude for all $N$.

In summary, we cannot here claim to have given the
complete solution to the problem of non-abelian field-theory
structure; but, rather, a new and complete functional
formulation (for eikonals and related self-energy graphs), and
a ``contiguity" method of extracting those terms which are
certainly going to be exponentiated, and which seem to
correspond to the identification of leading $\ln(E/m)$ dependence
appearing in the construction of specifically non-abelian
eikonals. It is hoped that these new techniques will be useful
for other processes, as discussed in the previous Sections.

\par  In particular, it is now appropriate to explain to the patient reader how
this procedure - which lacks manifest gauge invariance in a Yang-Mills context
- can be incorporated within a larger scheme, in order to obtain strictly
gauge-invariant results for physical scattering amplitudes.  There are three
separate issues involved.  In any eikonal calculation, one is searching for the
proper separation of longitudinal/timelike momenta from transverse momenta -
this is the problem attempted from first principles by Verlinde and Verlinde
\cite{VV 1}- while at the same time, one is trying to sum over the
contributions of all perturbative orders for the classes of graphs considered;
and, simultaneously, one must insist on the restrictions of gauge invariance.

\par The eikonal calculation of the present paper, with its ability to extract
leading-$ln(s)$ dependence, is intended to be used as an initial step in a
complete functional expression for the scattering of a pair of quarks, which
includes all gluonic self-interactions as part of a  ``gluonic sector"
described by the methods of Halpern \cite{Halpern}, or its slight
generalization by Fried \cite{HMF}.  The $A_{\mu}$-dependence of these
formulations takes the form of an exponential of linear and quadratic forms, so
that the $Q(s,t)$-propagator of (1.1) is now dependent upon auxiliary fields,
and is linked to subsequent functional integrals which describe the gluon
self-interactions; extra functional integrations maintain gauge restrictions.
The insertion of the forms of this paper then leads, as an intermediate step,
to a rather complicated set of functional integrals; but in the integrands of
these functional integrals, one has already extracted the leading $\ln (s)$
behavior of the simple eikonal where $s$ is essentially given by quark
kinematics.  For large $s$, by a rescaling of the auxiliary functional
integrands, one can now try to approximate and to extract relevant gluon
self-interaction structure, in this large $s$/small $t$ limit; and in a gauge
invariant way.  These calculations are presently underway, and whether they
will succeed is not yet known; but this is the reason why a functional
evaluation of the leading-log behavior of the simple eikonal form of (1.1) can
be relevant to quarks and gluons.

It is a pleasure to thank Jean Avan for some discussions
of doubly-ordered exponentials; Jean-Ren\'e Cudell for pointing
out expected group structures, in association with $\ ln(E/m)$
dependence, in models of quark-quark scattering; and T. T. Wu
for a critical discussion of the contiguity approach.

\end{document}